\documentclass{aa}
\usepackage{graphics,longtable}
\usepackage{graphicx}
\begin{document}

\title{LSS~1135: an O-Type Spectroscopic Binary in the galactic OB association Bochum~7}

\author{Mariela Corti\thanks{Fellow of CIC-BA, Argentina},
Virpi Niemela\thanks{Member of Carrera del Investigador Cient\'{\i}fico, CIC-BA, Argentina. Visiting Astronomer, CTIO, NOAO, operated by AURA, Inc., for NSF}
and Nidia Morrell\thanks{Member of Carrera del Investigador Cient\'{\i}fico, CONICET, Argentina}}

\institute{\scriptsize Facultad de Ciencias Astron\'omicas y Geof\'{\i}sicas,
       U.N. La Plata, Paseo del Bosque s/n, 1900 La Plata, Argentina}

\offprints{mariela@lilen.fcaglp.unlp.edu.ar}
\date{Received 24 February 2003 / Accepted 14 April 2003}
\authorrunning{Corti et al.}
\maketitle

\abstract{
From radial velocities derived from optical spectroscopic observations performed at CTIO, Chile, and CASLEO, Argentina, we have discovered that LSS~1135 is a single--lined O-type binary system with an orbital period of 2.7532 days. We present an analysis of the orbital elements of this system based on radial velocities of the He absorption lines. We classify the spectrum of LSS~1135 as O6.5V((f)). We also present spectral classifications and radial velocities for other seven OB stars in the region of Bochum~7, an OB association to which LSS~1135 belongs. Our data indicate a distance of 5.0~kpc for this star group. }

\maketitle

\section{Introduction}

 LSS~1135 ($\alpha_{2000}$ = 8$^h$ 43$^m$ 46$^s$ and $\delta_{2000}$ = -46$^{\circ}$ 07$^{'}$ 14$^{''}$) was assigned a spectral type OB and a photographic magnitude m$_{pg}$ = 11.2 in the catalog of Luminous Stars in The Southern Milky Way (LSS) (Stephenson \& Sanduleak). Moffat \& Vogt (1975) proposed from photoelectric photometry that LSS~1135 together with the Wolf-Rayet star LSS~1145 and other 7 OB stars listed in the LSS catalogue, form an OB association with galactic coordinates l = 265$^{\circ}$.20; b = -2$^{\circ}$.1. They called this star group Bochum~7 (Bo~7). 

Recently Sung et al. (1999) from UBVI CCD photo\-me\-try of the Bo~7 region have suggested that Bo~7 is part of the larger OB association Vela~OB~3.

 In this paper we present optical spectroscopic data of LSS~1135, which show that this star is a close binary system. We also present spectral types and radial velocities for the other 7 OB stars included by Moffat \& Vogt (1975) in Bo~7, namely LSS~1131, 1132, 1137, 1140, 1144, 1146, 1147.

 The paper is organized as follows: in Sect. 2 we describe the observations. In Sect. 3 we discuss the results and we present radial velocities and spectral classifications for stars in Bo~7, and  the orbital parameters for LSS~1135. In Sect. 4 we summarize our main results.
 
\section{Observations}

 The observational material consists of 21 photographic spectrograms and 29 digital CCD spectrograms of LSS~1135. We also obtained 13 photographic and 12 digital spectra of the other stars listed by Moffat \& Vogt (1975) as members of Bo~7. The instrumental configurations used are detailed in Table 1.

\subsection{Photographic spectrograms}

 The photographic spectrograms were all obtained by VSN between February 1982 and March 1985, at the Cerro Tololo Interamerican Observatory (CTIO), Chile. These spectra were secured with the Carnegie Image Tube Spectrograph (CITS)  attached to the 1m Yale reflector te\-les\-cope. All the exposures were made on Kodak III a-J emulsion  and widened to 1mm. A He-Ar lamp was used as comparison source.

 The photographic spectrograms were measured by VSN for the determination of radial velocities with an oscilloscope microdensitometer (GRANT engine) at Instituto de Astronom\'{\i}a y F\'{\i}sica del Espacio, Buenos Aires, Argentina. We also digitized some photographic spectrograms with a GRANT engine at La Plata Observatory, Argentina.

\subsection{Digital spectra}

 Digital spectral images of LSS~1135 were obtained between February 1997 and 1999, with the Boller \& Chivens (B \& C) and the REOSC Cassegrain echelle spectrographs attached to the 2.15 - m telescope at Complejo Astron\'omico El Leoncito (CASLEO)\footnote{Operated under agreement between CONICET, SeCyT, and the Universities of La Plata, C\'ordoba and San Juan, Argentina.} in San Juan, Argentina.

 Fourteen spectra of LSS~1135, and eleven of other members of Bo~7, were secured with the B \& C spectrograph using a PM 512$\times$512 pixels CCD detector with pixel size of 20$\mu$m. We used a 600 l~mm$^{-1}$ grating and the slit width was set to 200$\mu$.

 Fifteen spectra of LSS~1135 were obtained with the REOSC Cassegrain echelle spectrograph using as detector a TEK 1024x1024 pixels CCD, with pixel size of 24$\mu$m. We used a 400 l~mm$^{-1}$ grating as cross disperser and the slit width was set to 250$\mu$ and 300$\mu$. Ten \'echelle spectra of LSS~1135 were obtained in February 1998 binning the CCD by a factor 2.

 He-Ar (or Th-Ar with REOSC spectrograph) com\-pa\-rison arc images were observed at the same telescope position as the stellar images immediately after or before the stellar exposures. Also bias and flat-field frames were obtained every night, as well as spectra of flux and radial velocity standard stars.

 All digital, and digitized photographic spectra, were processed and analysed with IRAF\footnote{IRAF is distributed by NOAO, operated by AURA, Inc., under agreement with NSF.} routines at La Plata Observatory. Radial velocities were determined by fi\-tting Gaussian profiles to the spectral lines. Typical e\-rrors in our the radial velocities are approximately 20 km s$^{-1}$(standard error of the mean) for the instrumental configurations I and III, and 15 km s$^{-1}$   for the instrumental configuration II. The heliocentric radial velocities of the interestellar absorption lines measured in the high dispersion echelle spectra of LSS~1135 with their respective standard error are: Ca{\sc ii} K $\lambda$3933 \AA\ = 25 $\pm$ 5 km s$^{-1}$, Ca{\sc ii} H $\lambda$3968 \AA\ = 24 $\pm$ 4 km s$^{-1}$, Na{\sc i} $\lambda$5890 \AA\ = 25 $\pm$ 4 km s$^{-1}$ and Na{\sc i} $\lambda$5896 \AA\ = 27 $\pm$ 4 km s$^{-1}$. Only one component of each interstellar line is observed in our spectra.

\begin{center}
\begin{table*}
\caption{Instrumental configurations used}
\label{tab01}
\begin{tabular}{cllllcccc}
\hline
Nr.&Observatory&Epoch(s)& Telescope& Spectrograph& Recip.disp. & $\Delta\lambda$& exp.time& S/N\\
&       &          &  &             & (\AA~mm$^{-1}$) & (\AA)&  (min)& \\
\hline
I& CTIO  &1982 Feb. - 1985 March & 1-m   & CITS   & 45 & 3700-4900 &   20 &  30-60\\
II& CASLEO & (1997 - 1998) Feb. & 2.1-m & REOSC &  7 & 3700-6000 & 30 & 20-50 \\
III & CASLEO &1997 March - 1999 May & 2.1-m &B\&C & 115 &   3800-5000 & 20&  120-200 \\
\hline
\end{tabular}
\end{table*}
\end{center} 

\section{Results and their discussion}

\subsection{Spectral types and radial velocities of stars in the region of Bo~7}
 Our photographic spectrograms of stars listed as members of Bo~7 by Moffat \& Vogt (1975) at first glance confirm them as OB stars, as their spectra show absorption lines of hydrogen and helium. Approximate spectral types were determined initially by eye estimates of relative absorption line strengths in the photographic spectrograms, and then comparing the digitized and digital spectra with the digital spectral atlas of OB stars (Walborn \& Fitzpatrick, 1990). Spectral classifications are indicated in Fig. 1 depicting the spectra of stars observed in the region of Bo~7.

 For determination of radial velocities, we measured all absorption lines visible in the spectra. The radial velocity of LSS~1135 was found to be variable from night to night, and this star was therefore included in our program of studies of O type spectroscopic binaries. The results for LSS~1135 are presented separately below. Radial velocities for stars in the region of Bo~7 are listed in Table2.\\

\begin{center}
\begin{figure}
\includegraphics[angle=90, width=10cm]{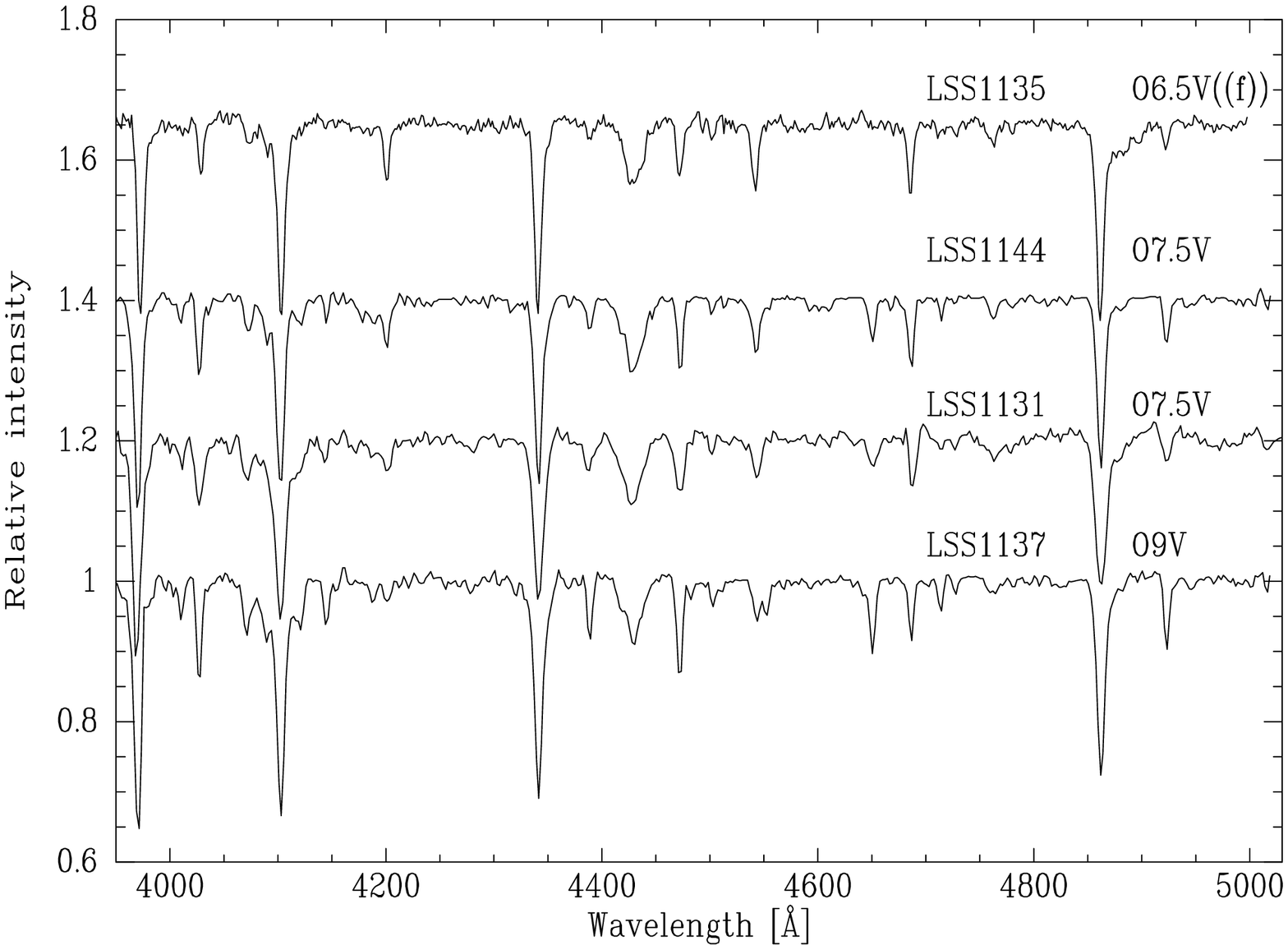}  
\includegraphics[angle=90, width=10cm]{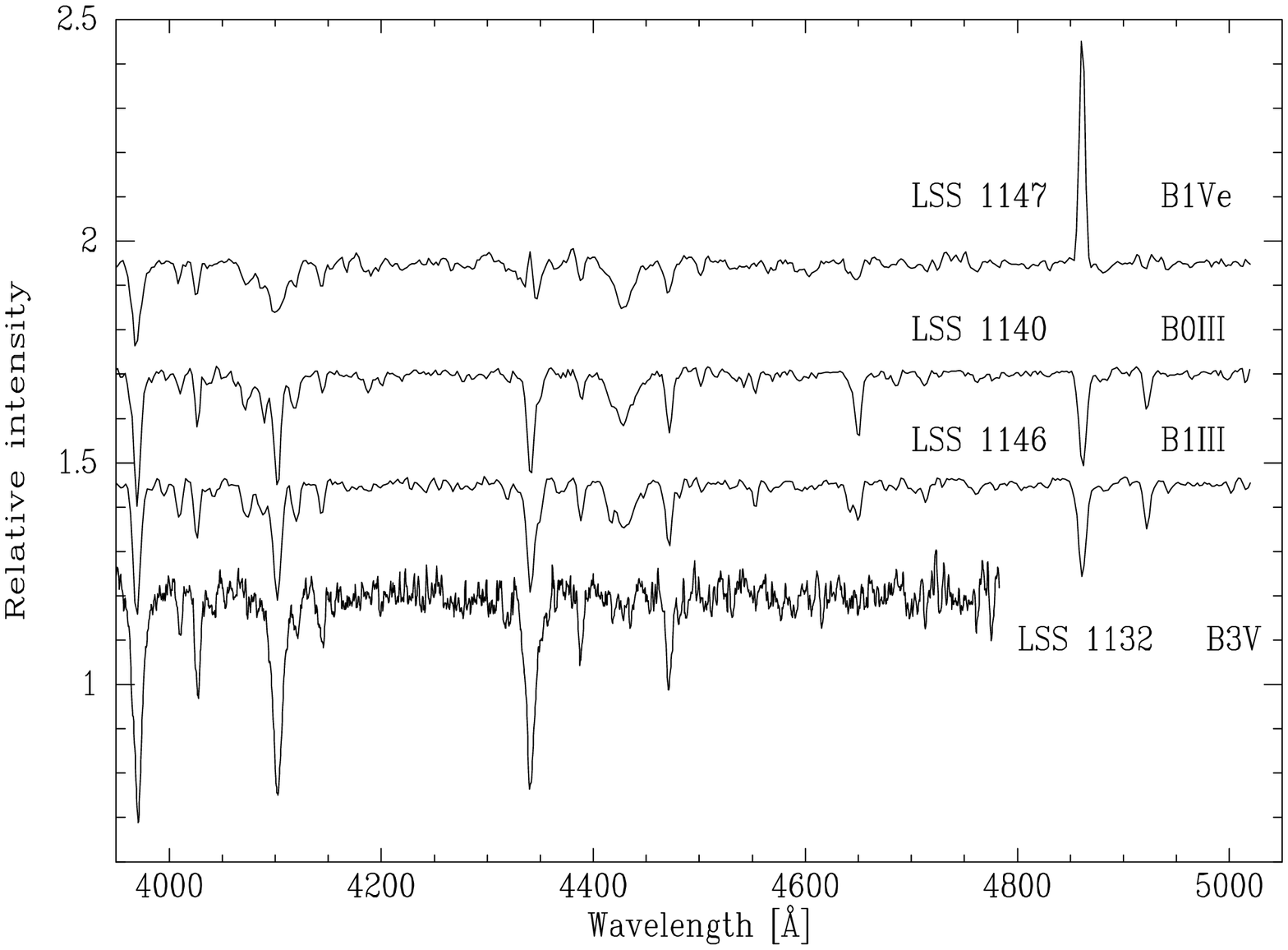}
\caption{Blue optical spectra of stars in the region of Bochum~7}
\label{fig01}
\end{figure}
\end{center}

\begin{center}
\begin{table}
\caption{Heliocentric Radial Velocities of Stars in the region of Bochum~7 }
\label{tab02}
\begin{tabular}{ccccl}
\hline
Star & HJD &\multicolumn{3}{c}{Heliocentric Radial Velocity }\\
\cline{3-5}
LSS & 2400000+  & IC & RV (km s$^{-1}$) & n \\ 
\hline
1131 & 45510.47 & I  & +64 &6\\
1131 & 45511.47 & I  & +90 &9\\
1131 & 51204.85 & III& +44 & 5 \\
1131 & 51219.58 & III& +61 & 6 \\
     &          &    &     &   \\
1132 & 45017.66 & I  & +03 &11\\ 
     &          &    &     &    \\
1137 & 45017.68 & I  & +56 &10\\
1137 & 51202.81 & III& +64 & 10 \\ 
1137 & 51204.83 & III& +55 & 10 \\
     &          &    &     &    \\
1140 & 45013.71 & I  & +73 & 10 \\
1140 & 45017.65 & I  & +40 & 6 \\ 
1140 & 51203.77 & III& +54 & 10 \\
1140 & 51219.83 & III& +58 & 8 \\
     &          &    &     &   \\
1144 & 45013.69 & I  & +74 & 11 \\ 
1144 & 45017.62 & I  & +78 & 7  \\
1144 & 45071.50 & I  & +65 & 8  \\
1144 & 45126.51 & I  & +57 & 8 \\ 
1144 & 51203.79 & III& +64 & 11\\
1144 & 51221.83 & III& +78 & 12 \\
     &          &    &     &   \\
1146 & 45509.48 & I  & +49 & 4 \\
1146 & 45512.47 & I  & +33 & 8 \\
1146 & 50868.66 & III & +48 & 11 \\
1146 & 51203.73 & III& +26 & 8 \\
1146 & 51220.84 & III& +64 & 8 \\
     &          &    &     &    \\
1147 & 45017.72 & I & +41 & 2 \\ 
1147 & 45071.61 & I & +55 & 2 \\ 
1147 & 50868.64 & III & +67 & 6 \\ 
\hline
\end{tabular}\\

\parbox{14cm}{{\scriptsize HJD $=$ Heliocentric Julian Date are in days.}}\\
\parbox{14cm}{{\scriptsize Nr. IC refers to the instrumental configurations listed in Table~1.}}\\
\parbox{14cm}{{\scriptsize n indicates the number of lines included in each mean value of the}}\\
\parbox{14cm}{{\scriptsize velocity.}}
\end{table}
\end{center}

\subsection{The spectrum of LSS~1135}
 Fig. 2 depicts one of the higher S/N digital spectra of LSS~1135 obtained at CASLEO with the instrumental configuration  III. This spectrum shows the absorption lines of He{\sc ii}, He{\sc i} and H characteristic of early O spectral type. In addition, several of our spectra show faint emission of N{\sc iii}
 $\lambda$4634--40~\AA\-. The diffuse interestellar band (DIB) $\lambda$4428 \AA\- and other interestellar absorptions are also present in the spectrum of LSS~1135, as well as in the spectra of other stars in Bo~7, except LSS~1132 which is a foreground object (cf. Fig1).

 For spectral classification of LSS~1135 in our medium resolution spectra we compared our spectra with the Digital Atlas of OB Stars published  by Walborn \& Fitzpatrick (1990) in the blue optical spectral region. This comparison yields a spectral type O6.5((f)). In our higher resolution spectra we also used the ratio of He{\sc i} $\lambda$5875 to He{\sc ii}$\lambda$5411 as described by Walborn (1980) for classification of O--type spectra in the yellow--red optical spectral region. In the blue spectral region of our high resolution spectra we used the equivalent width ratios of the
 quantitative spectral clasification for O-type stars of Conti \& Alschuler (1971) and Conti \& Frost (1977). All these confirm for LSS~1135 the spectral type O6.5V((f)).

 As an O type binary with a short period (see below), LSS~1135 appears as a potential candidate for showing the effects of colliding stellar winds. For this reason we searched for the Struve-Sahade (S-S) effect, which is produced by colliding stellar winds according to the model proposed by Gies et al. (1997). Examining our spectra in detail, we compared carefully one quadrature with the other in order to detect systematic differences in the relative strength of the He lines observed in the approaching and the receding phases. We did not detect in our data any spectral changes beyond the errors; therefore the S-S effect does not seem to be present in this binary system.  

The contribution of the secondary component to the spectral lines of LSS~1135 is not evident in our spectra, except perhaps in the Hydrogen Balmer lines whose radial velocity variations appeared of somewhat lower amplitude than those of He lines, when phased with the binary period (see below). We may therefore assume that the visual absolute magnitude of the secondary component is at least 2 magnitudes fainter than the O6.5((f)) star, which would correspond to an early B spectral type.

\subsection{The radial velocity orbit of LSS~1135}
 Table 3 lists the journal of observations and the mean heliocentric radial velocities, obtained averaging He{\sc i} $\lambda$4471, $\lambda$5875 and He{\sc ii} $\lambda$4199, $\lambda$4541 and $\lambda$4685 absorption lines in the spectra of LSS~1135. The numbers following the mean velocities of the absorption lines indicate how many lines were included in each average.

 As is evident from Table 3, the radial velocities of the absorption lines in the spectrum of LSS~1135 show large variations from one night to the other, but the radial velocities obtained during the same night do not show appreciable differences, implying a binary period of a few days. A period search algorithm (Marraco \& Muzzio, 1980) was applied to the radial velocities of Table 3.

 The best period found was P = 2.75318 $\pm$ 0.00002 days and no alias periods with similar probabilities were present. This period was entered as an initial value to calculate the orbital elements of LSS~1135. These were calculated with an improved version of the program initially published by Bertiau \& Grobben (1969). In the determination of the orbital elements we assigned weight 10 to the spectra obtained with the instrumental configuration II, and weight 1 for all others. The orbital elements are listed in Table 4. The errors quoted in this table are those calculated by the above mentioned program. The radial velocity variations and orbit of LSS~1135 are illustrated in Fig. 3.
\begin{center}
\begin{table}
\caption{Journal of observations of LSS~1135 }
\label{tab03}
\begin{tabular}{cccl|cccl}
\hline
HJD &\multicolumn{3}{c|}{Hel.R.V.}&HJD &
\multicolumn{3}{c}{Hel.R.V.}\\
\cline{2-4}\cline{6-8}
2400000+  & IC & He & n & 2400000+ & 
IC & He & n \\ 
\hline
45508.463 & I & 211 &4 & 50537.590 & III & -3 &4\\ 
45509.464 & I & 29 &3  & 50538.586 & III  & 170 &4\\
45510.456 & I & 21 &4  & 50540.596 & III & 14 &4\\ 
45511.453 & I & 187 &4 & 50541.567 & III & 160 &4\\
45512.452 & I & -37 &3 & 50542.553 & III & -46 &4\\
          &    &     &   &     & & & \\                          
45769.686 & I & 125 &4 & 50841.659 & II & 183 &4\\ 
45769.703 & I & 147 &3 & 50842.646 & II & -43 &5\\ 
45773.689 & I & 28 &4  & 50843.620 & II &  76 &5\\ 
45775.634 & I & 181 &4 & 50844.608 & II & 149 &5\\
          &    &     &   & 50845.620 & II & -46 &4\\ 
45842.494 & I & 45 &4  & 50846.661 & II & 165 &4\\ 
          &   &    &     & 50847.601 & II & 123 &3\\   
46132.582 & I & 38 &4  & 50848.609 & II & -52 &3\\
46132.724 & I & 2  &2  & 50849.612 & II & 189 &4\\ 
46133.661 & I & 150 &4 & 50851.584 & II &  0  &5\\
46134.569 & I & -23 &4 &              & & & \\                    
46134.639 & I & -2 &4  & 50854.700 & III & 59 &4\\ 
46135.647 & I & 57 &4  & 50855.608 & III & 122 &4\\
46135.739 & I & 103 &4  & 50858.601 & III & 102 &4\\
46136.639 & I & 123 &4 & 50859.600 & III & -36 &3 \\
46137.664 & I & -40 &4 & 50860.652 & III & 166 & 5\\
46138.663 & I & 121 &3 & 50861.721 & III & -14 &4\\
46139.773 & I & 17 &3  & & & & \\
        &&&              & 50963.442 & III & 4 &4\\  
50494.777 & II& 175 &4 & 50964.436 & III & -22 &4\\\
50495.794 & II& -38 &4 & 50965.447 & III & 182 &4\\
 & & & & & & & \\
50506.597 & II & 8  &4 & &&& \\
50507.542 & II & 50  &5& &&& \\
50508.549 & II & 192 &5 &   &  &  & \\
\hline
\end{tabular}\\

\parbox{14cm}{{\footnotesize{Notes: as in Table 2}}}
\end{table}
\end{center}

\begin{center}
\begin{figure} 
\includegraphics[angle=90, width=10cm]{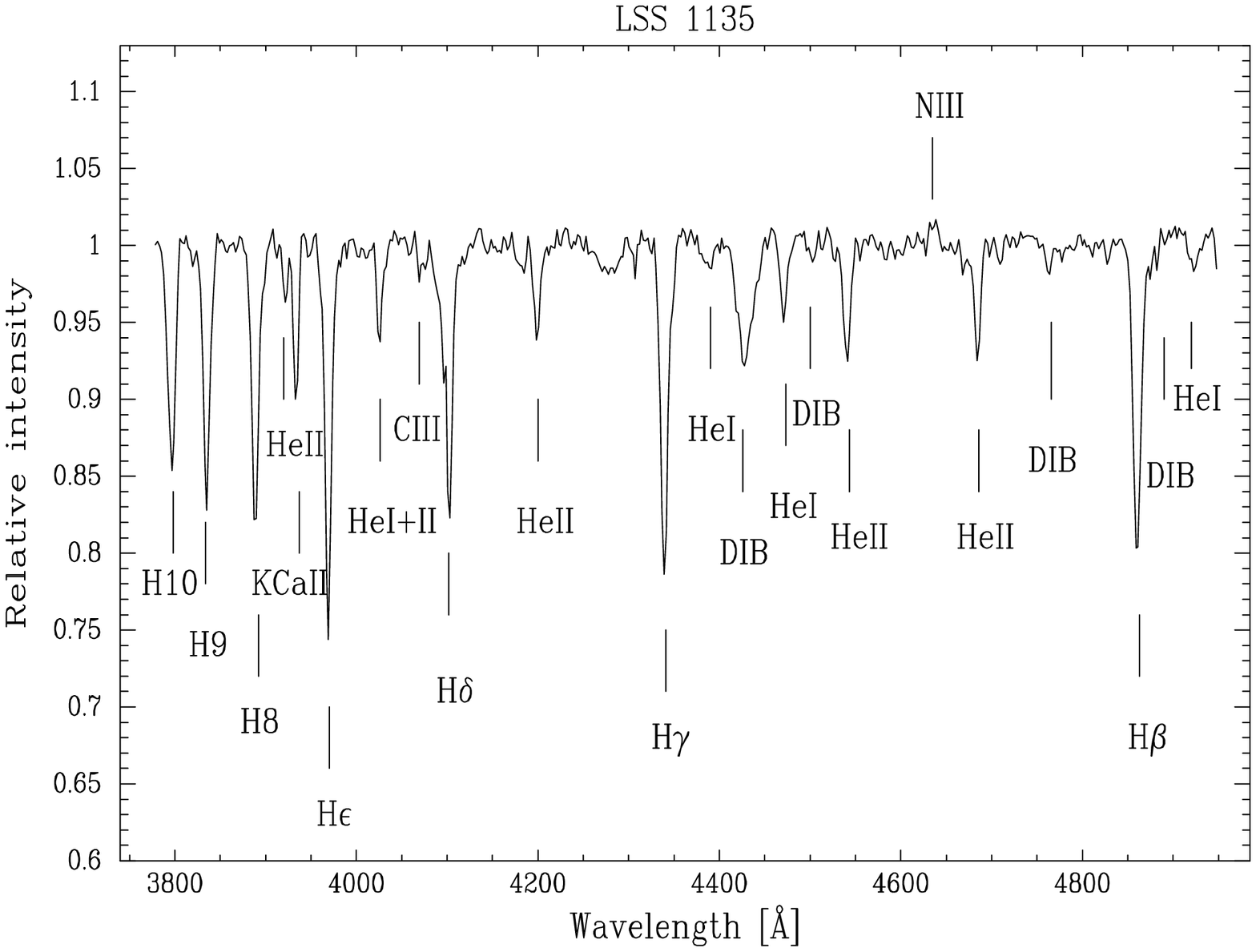} 
\caption{A continuum rectified spectrum of LSS~1135 obtained at CASLEO in April 1997 with the instrumental configuration III. Principal spectral features are identified.}
\label{fig02}
\end{figure}
\end{center}

\begin{center}
\begin{figure}
\includegraphics[angle=90, width=10cm]{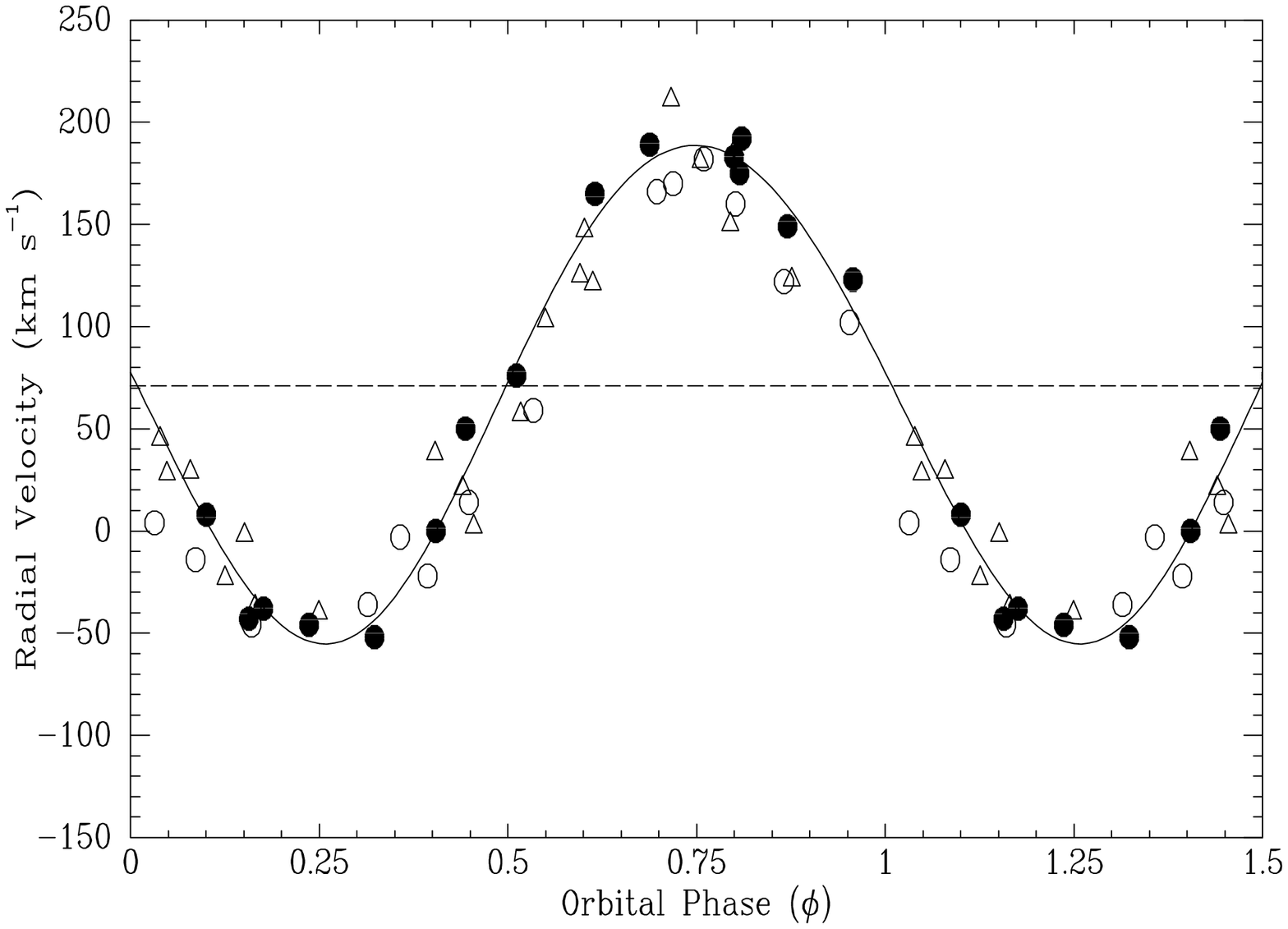}
\caption {Radial velocities of LSS~1135 from Table 3 phased with the period of 2.7532 days. Filled circles represent radial velocities measured in spectra observed with the instrumental configuration II, and empty triangles and empty circles those observed with the instrumental configuration I and III, respectively. Continuous curve represents the orbital solution from Table 4}
\label{fig03}
\end{figure}
\end{center}

\begin{center}
\begin{table}
\caption{Circular Orbital Elements of LSS~1135}
\label{tab04}
\vskip .2cm
\begin{tabular}{lccl}
\hline
 element   &  & &  \\
\hline
$a \sin i$  [R$_\odot$]  & 6.6 & $\pm$ & 0.11  \\
$K$ [km s$^{-1}$]       &  122 & $\pm$ & 2  \\
$V_{o}$ [km s$^{-1}$]    & 71 & $\pm$ & 1  \\
F(M) [M$_\odot$]        & 0.52 & $\pm$ & 0.03  \\
$T_{o}$ [HJD]           & 2450508.38 & $\pm$ & 0.01\\
$P$ [days]              & 2.75320 & $\pm$ & 1E$^{-5}$ \\
$\sigma$ [km s$^{-1}$]  &  9      &       &         \\ 
\hline
\end{tabular}
\end{table}
\end{center}
\subsection{Distance}
 In Table 5 we list data for the stars in the region of Bo~7. The 
succesive columns in this table indicate for each star: the number in the LSS catalogue (Stephenson \& Sanduleak, 1971); our estimate of the spectral type; V and B-V values from Moffat \& Vogt (1975); E(B-V) calculated from intrinsic colors corresponding to the spectral types; the spectrophotometric distance modulus determined adop\-ting the absolute magnitudes corresponding to the spectral types, according the calibration by Schmidt--Kaler (1982); stellar radial velocity (in km s$^{-1}$) referred to the LSR.

 The mean distance modulus for the stars in Bo~7 (~excluding the foreground object LSS~1132, and the emission line star LSS~1147) is 13.50, which corresponds to a distance of 5.0~kpc.
This is in good agreement with the distance found by Sung et al. (1999)
from CCD photometry. We also note that the photometric distance of LSS~1145 = WR~12 quoted by van der Hucht (2001), is 5 kpc, confirming that this star belongs to Bo~7 star group. The mean (LSR) radial velocities of the cluster members are between 55 and 40 km s$^{-1}$, (excluding LSS~1146, cf. note in Table 5) implying a kinematical distance between 5.0 and 6.0 kpc in circular galactic rotation model. Thus Bo~7  appears to be located at the extreme of the Perseus spiral arm in our
Galaxy (cf. Fig.5 in Russeil, (2003)).

\begin{center}
\begin{table}
\caption{Data for Stars in the Region of Bo~7}
\label{tab05}
\vskip .2cm
{\scriptsize
\begin{tabular}{llccccc}
\hline
LSS  & Sp.Type & V$^a$ & B-V$^a$ & E(B-V) & V$_0$-M$_v$ & RV\\
  &     &           &            &         &         & (LSR)  \\
\hline
1131      &     O7.5V & 10.80 & 0.51  & 0.83   & 13.33  & +50 \\
1132$^b$ & B3V  & 10.05 & 0.37  & 0.54   & 9.88   & -13 \\
1135      & O6.5V((f))& 10.88 & 0.40  & 0.73   & 13.95  & +55 \\
1137      &    O9V    & 11.38 & 0.49  & 0.80   & 13.40  & +42 \\
1140      &    B0III  & 11.65 & 0.76  & 1.05   & 13.49  & +40 \\
1144      &     O7.5V & 11.27 & 0.64  & 0.96   & 13.39  & +53 \\
1145$^c$  &       WN8 & 10.78 & 0.56  &        &        &  var.\\
1146$^d$  &     B1III & 11.57 & 0.50  &  0.82  & 13.43  & +26 \\
1147$^e$  &      B1Ve &  11.53& 0.58  &  0.78  & 13.11  & +54\\
\hline
\end{tabular}\\
}
\parbox{12cm}{{\scriptsize $a$ from Moffat \& Vogt (1975)}}\\
\parbox{12cm}{{\scriptsize $b$ foreground star, see also Lundstrom \& Stenholm (1984)}}\\
\parbox{12cm}{{\scriptsize $c$ WR 12 cf (van der Hucht, 2001). Spectroscopic binary (Niemela, 1982). }}\\
\parbox{12cm}{{\scriptsize $d$ Line widths appear variable. May be an unresolved }}\\
\parbox{12cm}{{\scriptsize double--lined binary?}}\\
\parbox{12cm}{{\scriptsize  $e$ H Balmer lines H$\beta$ and H$\gamma$ show central emission, see Fig. 1 }}\\

\end{table}
\end{center}
 
\section{Summary}

 A radial velocity study of the spectral lines of LSS~1135, a member of the galactic OB association Bo~7, shows that it is a single--lined O-type binary, which we classify as O6.5V((f)). We find an orbital period of 2.75320 days for this binary. The value of the semi-amplitude (K) of the radial velocity variations of  the He  lines is 122 km s$^{-1}$ and the mass function (F(M)) of the binary system is 0.52 M$_\odot$. The secondary component is not detected in our spectra, which means that it should be at least 2 magnitudes fainter, corresponding to an early B type star.

We also have observed 7 other OB stars in the region of Bo~7, for which we present spectral types and radial velocities. From these data we derive a spectroscopic and kinematical distance of about 5.0 kpc for Bo~7, and conclude that the Wolf-Rayet star LSS~1145~=~WR~12 is also member of this association.
\vskip 0.5cm
\noindent Acknowledgements
\vskip 0.3cm
\noindent We would like to thank the Directors and staff of CTIO and CASLEO for the use of their facilities. We also ack\-now\-ledge the use at CASLEO of CCD and data acquisition system partly financed by U.S. NSF Grant AST-90-15827 to Dr. R.M. Rich.\\
This research has received financial support from IALP, CONICET, Argentina. 
\vskip 0.5cm
\noindent References
\vskip 0.3cm 
\noindent Bertiau, F.C. \& Grobben, J., 1969, Ric. Astron. Sp. Vaticano 8, 1\\
Conti, P. \& Alschuler, W., 1971, ApJ 170, 325\\
Conti, P. \& Frost, S., 1977, ApJ 212, 728\\
Gies, D.R., Bagnuolo, W.G. \& Penny, L.R., 1997, ApJ 479, 408\\
Lundstrom, I. \& Stenholm, B., 1984, A\&AS 58, 163\\
Marraco, H. \& Muzzio, J.C., 1980, PASP 92, 700 \\
Moffat, A.F.J. \& Vogt, N., 1975, A\&AS 20, 85\\
Niemela, V.S., 1982, In Wolf-Rayet Stars: Observations, Physics, Evolution, Proc. IAU Sump. 99, eds. C. de Loore \& A.J. Willis, (Reidel - Dordrecht)\\
Russeil, D., 2003, A\&A 397, 133\\
Schaller, G., Schaerer, D., Meynet, G. \& Maeder, A., 1992, A\&AS 96, 269\\
Schmidt-Kaler, Th. 1982, In Landolt-Bornstein \& H.H. Voigt, (Springer-Verlag, Berlin)\\
Stephenson, C. \& Sanduleak, N., 1971, Publ. Warner and Swasey Obs. 1, 1\\
Sung, H., Bessell, M.S., Park, B.G. \& Kang, Y.H., 1999, JKAS 32, 109\\
van der Hucht, K.A., 2001, New Astronomy Review 45, 135\\
Walborn, N., 1980, ApJS 44, 535\\
Walborn, N. \& Fitzpatrick, E., 1990, PASP 102, 379\\

\end{document}